\documentclass[authoryear]{elsarticle}

\graphicspath{ {./figures/} }

\usepackage{multirow}

\usepackage{amsmath}

\newcommand{\fderiv}[2]{\ensuremath{\frac{\mathrm{d} #1}{\mathrm{d} #2}}}

\newcommand{\spscript}[2]{\mbox{#1}$^\mathrm{#2}$}

\journal{Biosystems}

\begin{document}
\def\floatpagepagefraction{0.95}
\def\textpagefraction{0.05}

\begin{frontmatter}

\title{Spatiotemporal bursting in simulated cultures of cortical neurons}

\author[1]{Michael Stiber\corref{cor1}}
\ead{stiber@uw.edu}
\ead[url]{faculty.washington.edu/stiber}

\author[1]{Natalie Gonzales}
\ead{ngonza5@uw.edu}

\author[2]{Jewel YunHsuan Lee}
\ead{jewel.yh.lee@gmail.com}

\affiliation[1]{organization={University of Washington Bothell, Computing and Software Systems Department},
  addressline={Box 358534, 18115 Campus Way NE}, city={Bothell},
  state={WA},postcode={98011}, country={USA}}

\affiliation[2]{organization={Ultrasound R\&D, Philips Healthcare}, addressline={22100 Bothell Everett Hwy}, city={Bothell}, state={WA}, postcode={98021}, country={USA}}

\cortext[cor1]{Corresponding author}
  
\begin{abstract}
  Cultures of neurons grown on multi-electrode arrays have
  become a common experimental preparation for investigating
  developing neural networks. Experiment and simulation have shown
  that these developing networks eventually exhibit bursting behavior
  in which the entire culture participates for short periods of time,
  with inter-burst intervals in which the network is comparatively
  quiescent. This paper extends previous simulation results by
  examining the spatiotemporal patterns of such bursting. We show that
  these bursts originate at a small number of network locations and
  propagate as waves of activity. We demonstrate that this type
  of activity does not require fine tuning of neuron or network
  parameters. We also examine how this activity changes during
  development and the dependence of such activity and its
  triggering on both local and global network properties.
\end{abstract}


\end{frontmatter}

\section{Introduction}

Neural circuits are nonlinear systems that produce an immense range of
complex behaviors. These circuits are composed of individual cells,
whose molecular and dynamical activities govern development of network
connectivity and give rise to an organism's sensory processing, motor
activity, and behavior (not to mention such phenomena as
consciousness). The relationships among these multiple spatiotemporal
levels of detail are complex, and the basic principles of how
individual neurons' activities produce development and computation in
networks is still largely unknown. One approach to examining this
question has been to study cultures of mammalian cortical neurons as
they form functioning networks on multi-electrode arrays (MEAs)
\citep{thomas_jr_miniature_1972,gross_simultaneous_1979,pine_recording_1980}. These
cultures allow stimulation and recording from individual cells, or
local groups of cells, over time as networks develop. As such, they
can serve as platforms for investigating the interplay among
individual neuron dynamics, network development, and network dynamics.

One of the major features of network dynamics in such cultures is a
convergence to whole-culture bursting
\citep{maeda_mechanisms_1995,vanpelt_long-term_2004,stegenga_analysis_2008,gritsun_growth_2012}. This
is a type of spontaneous activity in which most or all neurons in the
culture participate in synchronized spiking over a brief period of
time. These bursts begin to occur during network development,
repeating at irregular intervals and developing increased intensity
and decreased duration.

\citet{maeda_mechanisms_1995} demonstrated that these bursts originate
in small regions of the network and propagate directionally across
it. This was later confirmed in extensive experimental results in
which the originating sets of neurons were termed \emph{leaders}
\citep{eytan_dynamics_2006,eckmann_leader_2008,pasquale_leader_2013,lonardoni_recurrently_2017,pasquale_stimulation_2017}. Moreover,
these bursts were shown early on to propagate directionally across
networks \citep{maeda_mechanisms_1995}; later results have been
consistent with the interpretation of this as traveling waves of
activity \citep{sato_traveling_2012,okujeni_mesoscale_2017}, which is
consistent with observations of traveling waves in mammalian cortex
\citep{muller_cortical_2018,davis_spontaneous_2020,davis_spontaneous_2021,klavinskis-whiting_generation_2023}.

Simulations of such networks can exhibit many of the same behaviors.
\citet{gritsun_growth_2012} developed a detailed model of neurite
outgrowth in such preparations, and used that to produce connectivity
patterns that were then incorporated into simulations of network
spiking activity; data was collected from a simulated electrode
array. These showed whole-network bursting activity, inter-burst
timing, and wave-like propagation consistent with those observed in
living preparations. However, simpler models can produce similar
results; networks of spiking neurons and phase oscillators will
produce traveling waves originating at the equivalent of leaders
\citep{ermentrout_traveling_2001,zbinden_leader_2011,sakuma_simulation_2016,huang_propagation_2017,paraskevov_spatially_2017,senk_conditions_2020}.

The emergence of these spontaneous network bursts can be
contextualized within classical assembly theory, which posits that
cell assemblies---rather than individual neurons---serve as the
fundamental units of cortical circuits \citep{Palm_1982, Palm_2014,
  Yuste_2024}. Furthermore, the structured propagation of activity
(such as the traveling waves we will demonstrate here) is a
well-documented emergent property of recurrent networks, sharing
theoretical roots with models of synfire chains \citep{Abeles_1991,
  Zheng_2014, Miehl_2022, Cano_2026}.

In our own previous work
\citep{kawasaki_bursting_2012,kawasaki_simple_2014}, closed-loop,
MEA-scale simulations were developed that modeled the entire network
developmental period at the temporal resolution of individual neuron
spiking behavior. These simulations captured the effect of neural
activity on on neurite outgrowth, and vice versa, using simple neuron
and outgrowth models. Exploration of a set of model parameters showed
a broad region of that parameter space that produced stable network
bursting behavior. Examination of that behavior showed burst
development, burst shape, and inter-burst timing consistent with the
living preparations. This demonstrated that these aspects of network
dynamics could be substantially captured by closed-loop models with
relatively simple dynamics. It also reinforced the conclusion that
such network bursting was an emergent behavior that did not require
neuron models capable of autonomous bursting behavior.

This paper presents results from closed-loop, MEA-scale simulations of
development in cortical cultures that exhibit whole-network
bursting. It extends our previous work \citep{kawasaki_simple_2014},
which was limited to analyzing the global temporal statistics of
bursting (such as average network firing rates and inter-burst
intervals) due to the computational infeasibility at that time of
recording and processing individual spike histories. By capturing the
timing of every spike produced by every neuron over the entire
developmental period, we achieve the spike-level spatial resolution
necessary to transition from temporal to high-resolution
spatiotemporal analysis. This allows us to determine not just
\emph{when} bursts occur, but \emph{where} they originate, how they
propagate as traveling waves, and how the spatial distribution of
these origins condenses over development. Consequently, we can reveal
the mesoscale network influences that drive burst origination,
answering fundamental questions about the spatial organization of
spontaneous activity that cannot be addressed by global temporal
metrics alone.

\section{Modeling and Simulation}

The overarching goal of our work has been to develop models that
include the minimal dynamics at each level necessary to reproduce
behaviors seen in living systems, and then to simulate these models at
the time scale of spike production to capture the time of each spike
from every neuron. We utilized a closed-loop approach in which neural
activity drives structural plasticity (neurite outgrowth), which in
turn reshapes network activity. This aligns with biological evidence
that network activity and structure are co-dependent during
development \citep{van_ooyen_implications_1995}, and this continuous
feedback loop is necessary to observe emergent self-organization
without relying on external patterned input or more complex
single-neuron bursting mechanisms.  In brief, the model architecture
consists of a unified simulation where individual neurons exist on a
spatial grid, grow connectivity radii dynamically based on their
firing rates, and form synapses upon spatial overlap, establishing a
continuous feedback loop between structural development and electrical
activity. We have previously shown \citep{kawasaki_simple_2014} that
this model reproduces all major results observed in living MEA
preparations (whole-network bursting; evolution of burst temporal
shape---duration, onset synchronization, and intensity---during
development; inter-burst interval ranges).

The key model and simulation features, described in
subsequent sections, are:
\begin{itemize}
\item Leaky integrator neuron model, where each neuron is strictly
  excitatory or inhibitory, satisfying Dale's principle.
\item Spike-coupled network arranged in a 2D grid with a periodic
  tiling pattern to prevent random clustering artifacts.
\item Synapses that include dynamics of resource depletion and
  recovery.
\item Simple neuron-activity-dependent network development model
  mimicking calcium-regulated outgrowth, without requiring
  spike-timing-dependent plasticity.
\item Simulation of an equivalent 28 days of \textit{in vitro} network
  development with a 0.1ms time step.
\item Network behaviors as an emergent property; specifically,
  exploring variations in the balance of inhibitory and excitatory
  neurons and the null point for neurite outgrowth.
\end{itemize}

\subsection{Neuron and Synapse Models}

These simulations use a lumped, leaky integrator neuron model that
includes synaptic, bias, and noise currents
\citep{abbott_lapicques_1999}:
\begin{equation}
 C_m \fderiv{V_m}{t} = \frac{1}{R_m}(V_\mathrm{rest} - V_m) +
  I_\mathrm{syn} + I_\mathrm{inj} + I_\mathrm{noise}
  \label{eq:neuron}
\end{equation}
with $V_\mathrm{rest}$ being both the asymptotic and reset potential,
$I_\mathrm{syn}$ the total synaptic current, $I_\mathrm{inj}$ a
constant depolarizing current, $I_\mathrm{noise}$ a noise current, and
$C_m$ and $R_m$ the membrane capacitance and resistance,
respectively. When $V_m$ exceeds $V_\mathrm{thresh}$, the firing
threshold, a spike event is generated and $V_m$ is set to
$V_\mathrm{rest}$. This model also incorporates an absolute refractory
period, $T_\mathrm{refract}$. Parameters are set so that most neurons
do not produce spikes in the absence of excitatory input; a subset of
cells are selected as endogenously active and have their
$V_\mathrm{thresh}$ reduced so that they will spontaneously fire.

Generated spikes are transmitted to connected neurons after fixed
delays via synapses with activity-dependent facilitation and
depression \citep{markram_differential_1998,tsodyks_neural_1998} with
four state variables: three that govern the fraction of synaptic
resources in particular states --- $x$ (recovered state), $y$ (active
state), and $z$ (inactive state) --- and one, $u$, that represents
synaptic efficiency:
\begin{align}
  \fderiv{x}{t} &= \frac{z}{\tau_{\mathrm{rec}}} - u x
      \delta(t-t_{\mathrm{sp}}) \label{eq:x} \\
  \fderiv{y}{t} &= -\frac{y}{\tau_I} + U x \delta(t-t_{\mathrm{sp}}) \\
  \fderiv{z}{t} &= \frac{y}{\tau_I} - \frac{z}{\tau_{\mathrm{rec}}} \\
  \fderiv{u}{t} &= -\frac{u}{\tau_{\mathrm{facil}}} + U(1-u)
      \delta(t-t_{\mathrm{sp}})
  \label{eq:u}
\end{align}
Here, $\delta(t-t_\mathrm{sp})$ is the unit impulse at the arriving
spike time $t_\mathrm{sp}$. The three time constants $\tau_I$,
$\tau_\mathrm{rec}$, and $\tau_\mathrm{facil}$ govern inactivation
after an arriving spike, recovery from inactivation, and facilitation
after a spike, respectively. The synaptic current produced by an
arriving spike is $I_\mathrm{syn}=Wy$, where $W$ is the strength of
the connection determined by the development model, described in the
next section. Synapse parameters vary according to the combination
of presynaptic and postsynaptic neuron type (excitatory or inhibitory)
\citep{kawasaki_simple_2014}.

\subsection{Development and Network Models}

The cortical culture model is composed of a $100 \times 100$ array of
neurons, arranged in a rectangular grid as a $10 \times 10$ pattern of
inhibitory, excitatory, and endogenously active cells that is tiled 10
times horizontally and vertically to form the whole network
\citep{kawasaki_simple_2014}. This regular tiling pattern, consistent
with our principle of model simplicity, avoids chance clustering that
might bias network activity; the random alteration of endogenously
active neurons' firing thresholds introduces variability across the
network.  Existence and strength, $W$, of connections between any two
neurons is governed by a neurite outgrowth model in which each
neuron's set of neurites is modeled as filling a circle centered on
that neuron with uniform density. The radius of each neuron's circle
of connectivity varies according to a phenomenological rule developed
to reflect regulation by intracellular calcium concentration
\citep{van_ooyen_implications_1995,van_ooyen_complex_1996}:
\begin{align}
  \fderiv{R_i}{t} &= \rho G(F_i) \label{eq:outgrowth} \\
  G(F_i) &= 1 - \frac{2}{1 + \exp((\epsilon - F_i )/\beta)} \label{eq:g}
\end{align}
Here, $R_i$ is the radius of neuron $i$'s circle of connectivity,
$F_i$ is neuron $i$'s recent average firing rate (normalized within
the range $[0, 1]$), $\rho$ is a rate constant, $\epsilon$ is the
outgrowth null point, and $\beta$ controls the sensitivity of
outgrowth to $F_i$.

Connections are established between neurons that have circles of
connectivity that overlap, with $W$ for them proportional to area of
overlap, consistent with the simplifying assumption that neurite
density is uniform within such circles and thus number of synapses
will be proportional to area of overlap.

In this model, each neuron determines its radius of connectivity by
its own spiking behavior, which for all neurons that are not
endogenously active is a result of the net excitation it receives from
nearby, connected, neurons. This feedback effect motivates the
closed-loop nature of our simulations. Differing neuron
characteristics, both inherent and as a result of their location in
the culture, produce variations in activity and connectivity
\citep{kawasaki_simple_2014}. For example, neurons at or near the
edges of the culture have fewer neighbors within any given radius; as
a result, they generally grow larger radii of connectivity to receive
the same net stimulation and thereby spike at the null rate,
$\epsilon$.

\subsection{Simulation and Data Collection}
\label{sec:sim-data}

We used the Graphitti graph-based systems simulator
\citep{stiber_uwb-biocomputinggraphitti_2025,stiber_uwb-biocomputingbraingrid_2018,stiber_braingridworkbench_2017}
to perform simulations equivalent to 28 days' development for four
networks with two different values each for the growth parameter
$\epsilon$ (1.0 and 1.9 spikes/sec) and the network structure
parameter governing fraction of cells that were excitatory (90\% and
98\%). These specific parameter values were chosen because previous
parameter space explorations identified them as points that yield
stable, stationary bursting \citep{kawasaki_simple_2014} and thus we
repeat those specific simulations with identical parameters to provide
a direct comparison.  Selected neuron and synapse parameter values are
summarized in Table~\ref{tab:params}; complete lists are provided in
\citet{kawasaki_simple_2014}.  The biological time scale mapping (to
28 Days \textit{in vitro}) relies on a 40x computational acceleration
applied only to the outgrowth rate $\rho$ (shown in the table as its
reciprocal, the growth time constant). The membrane time constant is
separated by a factor of roughly $10^5$ from this accelerated growth
time constant, and further increases to the outgrowth rate were
computationally verified to have no impact on neuron spiking behavior.

\begin{table}
\centering
\caption{Key neuron and synapse parameter values used in simulations.}
\label{tab:params}
\begin{tabular}{|l|l|p{1.5in}|}
\hline
\textbf{Parameter} & \textbf{Summary} & \textbf{Value} \\ \hline
  $t_m$ & Membrane time constant & 30 ms \\
  $1/\rho$ & Growth time constant & 10,000 sec \\
$V_\mathrm{rest}$ & Resting membrane voltage & 0.0 V \\
$V_\mathrm{thresh}$ & Threshold voltage & 15.0 mV (excitatory/inhibitory) \\
 &  & [13.565, 13.655] mV (spontaneous) \\
$V_\mathrm{reset}$ & Reset voltage & 13.5 mV \\
$T_\mathrm{refract}$ & Absolute refractory period & 3.0 ms
                                                    (excitatory) \\
  & & 2.0 ms (inhibitory) \\ \hline
\end{tabular}
\end{table}

Simulations were run on an Intel Xeon server running Ubuntu Linux
16.04.3 using an NVIDIA Tesla K80 GPU with CUDA 8.0 libraries. Spike
times (and corresponding neuron $(x,y)$ location) were recorded to
HDF5 files with 0.1ms resolution, along with total spike counts for
the entire network within 10ms bins. Binned spike counts for each
simulation were compared to previous results
\citep{kawasaki_simple_2014} (in which such counts were the basis for
all analysis, consistent with data collected from the majority of MEA
data) to verify that the new and old simulation results were
qualitatively identical (unavoidable differences in noise realizations
prevent precise numerical equality).

\section{Analysis}
\label{sec:burstMethods}

Seven major categories of data analyses were performed, as described in
detail in the following subsections:
\begin{enumerate}
\item Whole-network bursts, and their constituent spikes, were
  identified and their spatiotemporal patterns
  were visualized.
\item Burst origin locations were determined.
\item Bursts that originated independently near the same time, and
  that therefore produced overlapping wavefronts, were identified and
  excluded from further analysis.
\item The speed of propagation of the burst wavefronts was computed
  and the variation in mean propagation speed during network
  development was visualized.
\item The sequence of burst origin locations during network
  development was described and their spatial distribution and
  sequential correlation were analyzed.
\item The relationship during development between burst origins and
  endogenously active neuron spiking rate was investigated.
\item The relationship between burst origin and distance to network
  edge (edge effects) was examined.
\end{enumerate}
Analyses were performed in Matlab; code and data are available on Zenodo
\citep{stiber_uwb-biocomputinggraphsystemsanalysis_2026,stiber_2026_21048146}.

\subsection{Burst Identification and Visualization}

Bursts were identified using the count of spikes produced by the
entire network within 10ms bins. An APNFR (Average Per-Neuron Firing
Rate) threshold method was chosen for consistency with methods in the
physiological literature regarding bursting in cortical cultures.
Sequences of bins with spike counts above 50 (corresponding to a spike
rate threshold of 0.5 spikes/second/neuron) were considered to be
within bursts. This approach parallels that used in the MEA
literature; the specific threshold value was chosen by
\citet{kawasaki_simple_2014} through analysis of APNFR statistics to
reliably separate bursting from background firing. The first bin in a
sequence marked the burst start and the final bin marked the burst
end, providing 10ms accuracy for these events. These start and end
times were then used to select, for each burst, the set of spikes
produced by each neuron that were considered part of the
burst. Because there was ongoing background neuronal activity, a very
small number of spikes associated with each burst were not causally
connected to the bursting behavior itself, but they were
insignificantly few compared to the number of spikes in each burst.

These spikes were then used to visualize each burst's spatiotemporal
pattern. Bursts were segmented into 10 ms non-overlapping windows and
the number of spikes produced by each neuron in the $100 \times 100$
network was used to generate a heatmap image representing spiking
intensity with color (blue to red) indicating spiking rate (zero to
highest in that window).  These images were either output individually
or concatenated to produce videos.

\subsection{Burst Origins}\label{sec:origins}

Previous analysis \citep{kawasaki_simple_2014} was based solely on
binned spike counts and thus only examined bursts as temporal
phenomena. Examination of spatiotemporal patterns here revealed that
every burst was invariably initiated within a very small, compact
cluster of cells (as can be seen in figure~\ref{fig:burst-wavefront}).
Each burst $b_i$ initiation was therefore analyzed to assign an
$(x_i,y_i)$ \emph{origin} point. The 10ms windows of burst
visualization were used for this purpose. We scanned these windows
from the start of each burst to identify the first one containing
sufficient spiking to clearly separate compact burst activity from
background activity.  This was done by selecting the first window in
which at least three neurons produced at least two spikes each. We
empirically verified that this was the minimum threshold that reliably
produced a clear separation of the start of bursting from background
activity that allowed identification of a burst origin (i.e., any
lower number of neurons or spikes produced obviously wrong origins;
raising these thresholds only resulted in identifying later windows).

Within that first window, those neurons that produced the most spikes
were identified as the set of most active cells. To guard against
multiple bursts originating near each other in time (and therefore
multiple origins; from the current data, less than one in 5000
bursts), we computed the distances between each pair of neurons in
that set. If there were multiple origins, this would correspond to
multiple subsets and neurons would have bimodal distributions of
distances (distances within their subset and distances between the
subsets). We discarded those neurons comprising the smaller subset. We
also manually examined each of the small number of associated bursts
to double-check that there were no unreasonable artifacts. This was an
initial filter for temporally overlapping bursts; as described next,
we also implemented a secondary filter and performed a second manual
check. The burst origin $(x_i,y_i)$ was then computed as the centroid
of the set of remaining most active cells.

\subsection{Overlapping Burst Identification}
\label{sec:overlappingBursts}

Bursts that originated near each other in time produced overlapping
wavefronts that precluded further analysis for the purposes of this
investigation. A simple method was implemented, similar to computation
of burst wavefront described in section~\ref{sec:propSpeed}, to
identify and exclude such overlapping bursts. In each 10 ms window in
each identified burst, the set of neurons that produced the most
spikes were identified; these were on the expanding burst
wavefront. Because the method of section~\ref{sec:origins} only
generated one origin point, if there were overlapping bursts, there
would be two subsets: those on a wavefront centered on that origin and
those centered on some other point. If there were at least three
points in the set (the minimum to allow comparison), their distances
to the burst origin were computed, essentially computing a Hough
transform. If there was a single burst, these distance values will all
be very similar, as the wavefront would be roughly a circular arc (or
set of circular arcs). If there were multiple bursts, then there would
be greater variability in distances.

The standard deviation of the distances was used as the metric for
determining if there was sufficient variability to diagnose multiple
bursts. We compared the standard deviation was compared to a threshold
(a distance of 5.0 was used) and, if greater than that, the data for
this ``burst'' was considered to include overlap from multiple bursts
and excluded from further analysis. Since a burst's data was excluded
if any of its 10 ms windows exceeded this threshold, we chose it
conservatively and manually examined every excluded burst to guard
against false positives. We then spot-checked a random 5\% of all
bursts from all simulations to guard against false negatives.

\subsection{Burst Propagation Speed}\label{sec:propSpeed}

As bursts invariably originated at a single point and then propagated
as wavefronts, the wavefront propagation speed was then determined for
each burst to see if this changed during network development. This
analysis was performed for each window from 100 ms after burst start
until 20 ms before burst end. These were chosen to focus on windows
where the most active neurons would lie on a substantial circular
wavefront arc, rather than a central cluster or the burst's
``remnants'' near the edge of the culture.  For each window, those
neurons that produced the most spikes were identified, their distances
to the origin were determined (from previous filtering, we were
assured that all of these distances were similar and thus on a
circular arc), the average distance was computed, and that value and
the time from burst initiation was used to compute the propagation
speed for that window. The computed speeds for each window making up a
burst were then averaged to produce an average propagation speed for
that burst. As the spatial grid used in the simulation is
dimensionless, speed is presented in units of \spscript{ms}{-1}.

\subsection{Burst Origin Sequences}

The temporal sequence of burst origins $(x_i,y_i)$, $1 \leq i \leq B$
(where $B$ was the total number of non-overlapping bursts in a
simulation), was then examined. These were first visualized by
segmenting the entire simulation duration into 25 non-overlapping
temporal windows, to identify if origin spatial pattern changed during
network development. For each window, the origins were plotted in the
2D plane with a line segment connecting each $(x_i,y_i)$ with
$(x_{i+1},y_{i+1})$ to create \emph{basic burst origin sequence}
plots. Combining these plots together into a single $5 \times 5$
display created a simple visual representation of how burst origin
patterns changed as network connectivity developed.

A variety of techniques were used to discern if the spatial sequence
of burst origins within the network exhibited behavior that could be
distinguished from random. Since the sequence pattern clearly changed
as the network developed, this analysis was performed on the same 25
windows as above. Origin locations were categorized by dividing the
$100 \times 100$ network into $10 \times 10$ tiles, corresponding to
the simulation tiling pattern of excitatory, inhibitory, and
endogenously active cells. Each origin was then assigned to the tile
it fell within, so each $(x_i,y_i)$, $1 \leq x_i, y_i \leq 100$, was
replaced with $(x'_i,y'_i)$, $1 \leq x'_i, y'_i \leq 10$; these were
numbered sequentially as $\ell_i$, $1 \leq \ell_i \leq 100$ (for the
100 tiles). The count of origins for each tile was used to visualize
the frequency of origins for each tile. Since origins tended to
cluster into a small number of tiles, only tiles that had non-zero
origin counts were plotted as \emph{burst origin histograms}. These
histograms were tested against uniform distributions using a $\chi^2$
test with $p<0.05$ indicating non-uniform distributions.

These histograms were used to assess whether all origin clusters
were equally likely. We also investigated sequential correlations
between bursts by computing the joint frequencies of $(\ell_i,
\ell_{i+1})$, which we plotted as $5 \times 5$ displays of 2D
images. As with simple histograms, only nonzero $\ell$ values were
used in this analysis.

\subsection{Burst Origins and Endogenously Active Neurons}
\label{sec:origin-active-methods}

All activity within these networks was driven by the endogenously
active neurons; as described by \citet{kawasaki_simple_2014}, these
were excitatory neurons with lowered spiking thresholds, thus capable
of producing spikes solely due to their inherent noise currents. It is
therefore reasonable to posit that bursts might be more likely to be
triggered in regions where there is greater excitatory drive from
these cells.

To test this hypothesis, we examined the correlation between burst
origin location and local excitatory drive.  Each $10 \times 10$ tile
established a regular pattern for the location of endogenously active
neurons, which was done to avoid the inevitable spatial clustering
that would have resulted if they were randomly distributed with
independent probability. In these simulations, 10\% of neurons (1000
out of the 10,000 total) were endogenously active. Each of these
neurons was assigned a lowered threshold chosen randomly from a
uniform distribution over a range that produced biologically realistic
spiking rates.

The average spiking rate of an endogenously active neuron was
therefore related to its lowered spiking threshold, and so the
difference between this threshold and that of non-endogenously-active
cells is a proxy for this average rate, and therefore its contribution
to excitatory drive, \emph{in the absence of synaptic input}. We then
related this location-based endogenous drive for the network with
burst origin location by averaging the lowered thresholds of the
endogenously active neurons within each $10 \times 10$ tile and
plotting that as an image with the burst origins overlaid. We also
computed Spearman's $\rho$ and corresponding $p$ for each simulation
between number of origins in each tile and average threshold reduction
for each tile.

\subsection{Edge Effects}

To assess whether the finite size of the simulated network introduced
biases in burst initiation locations, the spatial distribution of
burst origins relative to the culture edges was analyzed. For each
non-overlapping burst, the minimum distance from its origin coordinate
$(x,y)$ to the nearest edge of the $100 \times 100$ neuron grid was
computed. Because the number of neurons at a given distance from an
edge decreases towards the center of the culture, the raw frequency of
burst origins at each distance does not directly indicate spatial
bias. To account for this, the network was divided into concentric,
square-shaped bin strips based on distance from the edge. The count of
burst origins falling within each bin was then normalized by the area
of that bin to compute origin density. This density was examined as a
function of distance from the nearest edge to determine if burst
initiation was disproportionately favored at any distance from network
boundaries.

\section{Results}\label{sec:results}

\begin{table}
  \caption{Basic burst statistics for all four simulations.}
  \label{tab:basicBurstStats}
  \centering
  \begin{tabular}{|c|c|c|} \hline
    $(\epsilon, f)$ & \textbf{Total Bursts} & \textbf{Overlapping Bursts} \\ \hline
    $(1.9, 0.90)$ & 9,729 & 238 \\
    $(1.9, 0.98)$ & 9,160 & 126 \\
    $(1.0, 0.90)$ & 18,473 & 3,593 \\
    $(1.0, 0.98)$ & 17,659 & 844 \\
    \hline
  \end{tabular}
\end{table}

As mentioned in section~\ref{sec:sim-data}, previous work had
determined that sustained bursting with stationary neuronal outgrowth
occurred for the four simulations whose results are presented
below. In all four cases, during development neurons grew their radii
of connectivity for a short time in which no bursts occurred;
bursting began shortly after connections were formed between nearest
neighbors. These bursts' statistics changed markedly as the
simulations developed toward stationarity. Results in this paper are
from analysis of all bursts from the four simulations
$(\epsilon, f) = \{1.0, 1.9\} \times \{0.90, 0.98\}$, where $\epsilon$
is the outgrowth null point (firing rate at which a neuron's radius of
connectivity would not change) and $f$ was the fraction of the neurons
that were excitatory (and thus $1-f$ was the fraction that were
inhibitory). Basic statistics are presented in table~\ref{tab:basicBurstStats}.

\subsection{Burst Spatiotemporal Patterns}

\begin{figure*}
  \centering
  \includegraphics[width=\textwidth]{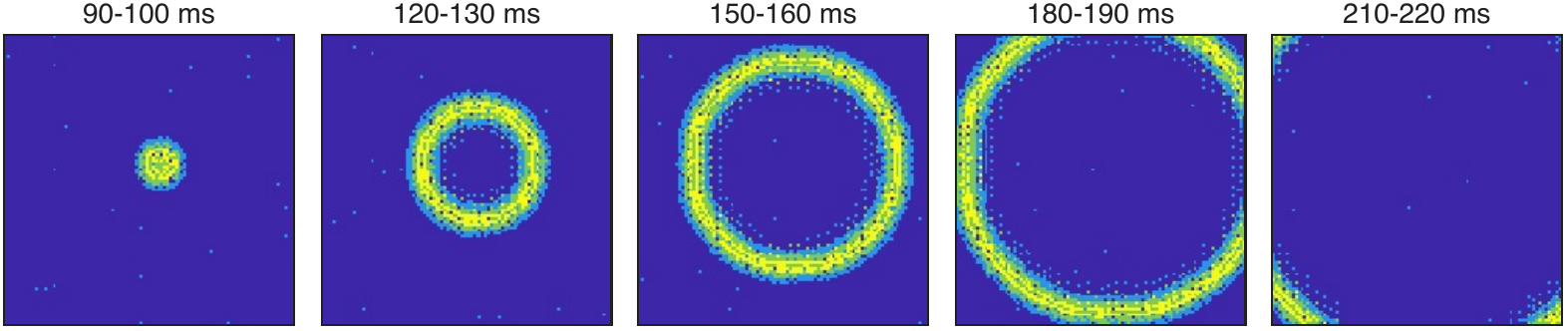}
  \caption{Example of a typical whole-network burst (chosen with
    origin near center of network for clarity). Each image includes
    10ms of activity; lighter color corresponds to larger number of
    spikes by that particular neuron (dark blue corresponds to no
    spikes).}
  \label{fig:burst-wavefront}
\end{figure*}

Figure~\ref{fig:burst-wavefront} presents an example visualization of
a typical whole network burst. Previous results indicated that bursts'
durations and peak amplitudes changed markedly as network development
progressed and that they originated at irregular time intervals that
were indistinguishable from random \citep[i.e., their inter-burst
intervals follow a distribution characteristic of a memoryless Poisson
or renewal process with dead time, lacking any deterministic or
periodic structure;][]{kawasaki_simple_2014}. The spatiotemporal
pattern results presented here invariably showed bursts originating at
a single location and spreading as a wave of activity across the
network. Occasionally, more than one burst would originate near the
same time; these were excluded from further analysis as described in
section~\ref{sec:overlappingBursts}; the number of those overlapping
bursts is presented in table~\ref{tab:basicBurstStats}.  The figure
also shows the scattered, isolated spiking of endogenously active
cells.


\begin{figure*}
  \centering
  \includegraphics[width=\textwidth]{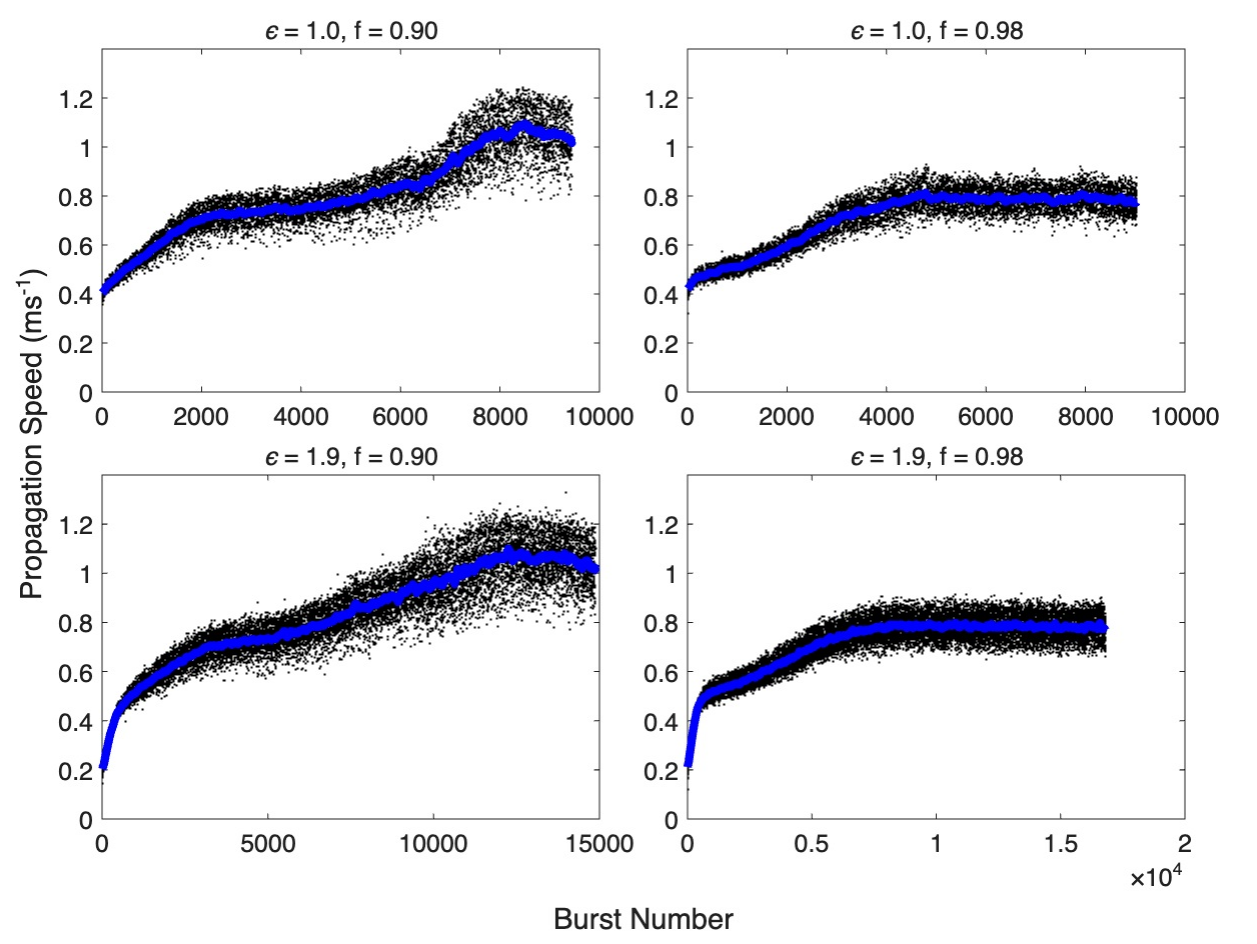}
  \caption{Burst speed during network development for individual
    bursts (points) and moving average of 100 bursts (bold line) and
    for given $\epsilon$ and $f$. While burst speed varies somewhat
    among bursts close in time, there is a clear trend in all cases.}
  \label{fig:burstSpeed}
\end{figure*}

Given previous results' indication that bursts became briefer and more
intense during network development, while still being composed of
roughly the same number of spikes, and given these new results that
from the start these bursts were waves of activity, we concluded that
these developmental changes were likely due to increasing burst
propagation speed. We analyzed burst wave propagation speed as a
function of time during development as shown in
figure~\ref{fig:burstSpeed}. These graphs display propagation speed of
individual bursts, along with a 100-burst moving average, versus burst
number. Results are consistent with previous observations for all four
bursting simulations in that faster wavefront propagation corresponds
to briefer burst duration. The figure reveals a rapid increase in
propagation speed early in development that levels out near simulation
end (as the preparation's development becomes stationary). Speeds
start around 0.2\spscript{ms}{-1} (for $\epsilon=1.9$) to
0.3\spscript{ms}{-1} (for $\epsilon=1.0$), meaning that it takes
around 3.3--5.0ms to propagate from one neuron to a neighbor
(horizontally or vertically, since these simulations used rectangular
connectivity; diagonal propagation would be $\sqrt{2}$ times
greater). Speeds settle to around 0.4 (for $f=0.98$) to
0.5\spscript{ms}{-1} (for $f=0.90$), or 2--2.5ms to propagate one
neuron. This compares to synaptic transmission delays used in the
simulation ranging from 0.8ms to 1.5ms.


\begin{table}
  \caption{Spearman's rank correlation computed to assess the
    relationship between burst speed and burst origin.}
  \label{tab:speedLocCorrel}
  \centering
  \begin{tabular}{|l|l|l|}\hline
    $(\epsilon, f)$ & $\rho$ & $p$ \\ \hline
    $(1.9, 0.90)$ & 0.037 & 0.000006 \\
    $(1.9, 0.98)$ & 0.11 & 0.0 \\
    $(1.0, 0.90)$ & -0.069 & 0.0 \\
    $(1.0, 0.98)$ & 0.017 & 0.116 \\
    \hline
  \end{tabular}
\end{table}

Because there is a fair amount of variation in the individual bursts'
wave propagation speed during each stage of development, we tested the
hypothesis that this is correlated with burst origin. This might be
reasonable to expect, as the path of wavefront propagation depends on
the origin location and different paths could correspond to sequences
of synapses with different strengths. We performed a quick test to see
if origin location was correlated with burst speed by converting each
burst origin location $(x,y)$ to linear index in column-major order
$i = 100y +x$ and computing Spearman's rank correlation between burst
speed and $i$, shown in table~\ref{tab:speedLocCorrel}. One can see
that, in all cases, correlations are either negligible to very low or
not statistically significant. Thus, we have in some (but not all)
cases evidence for trivial (but not practically meaningful)
association between burst origin and speed.

\subsection{Burst Origins}

To ensure conceptual precision, we define a burst \emph{origin} as the
$(x, y)$ spatial location where a network burst initiates. In the
literature, highly active individual neurons that frequently lead
bursts are termed Major Burst Leaders
\citep[MBLs,][]{pasquale_leader_2013}. Because our spatial resolution
allows us to see that burst initiation is a regional phenomenon, we
refer to these localized initiation zones as functional communities
\citep[fCOMs,][]{lonardoni_recurrently_2017}. Thus, an fCOM is a
spatial cluster at an $(x, y)$ origin location structurally composed
of individual MBLs.

\begin{figure*}
  \centering
  \includegraphics[width=\textwidth]{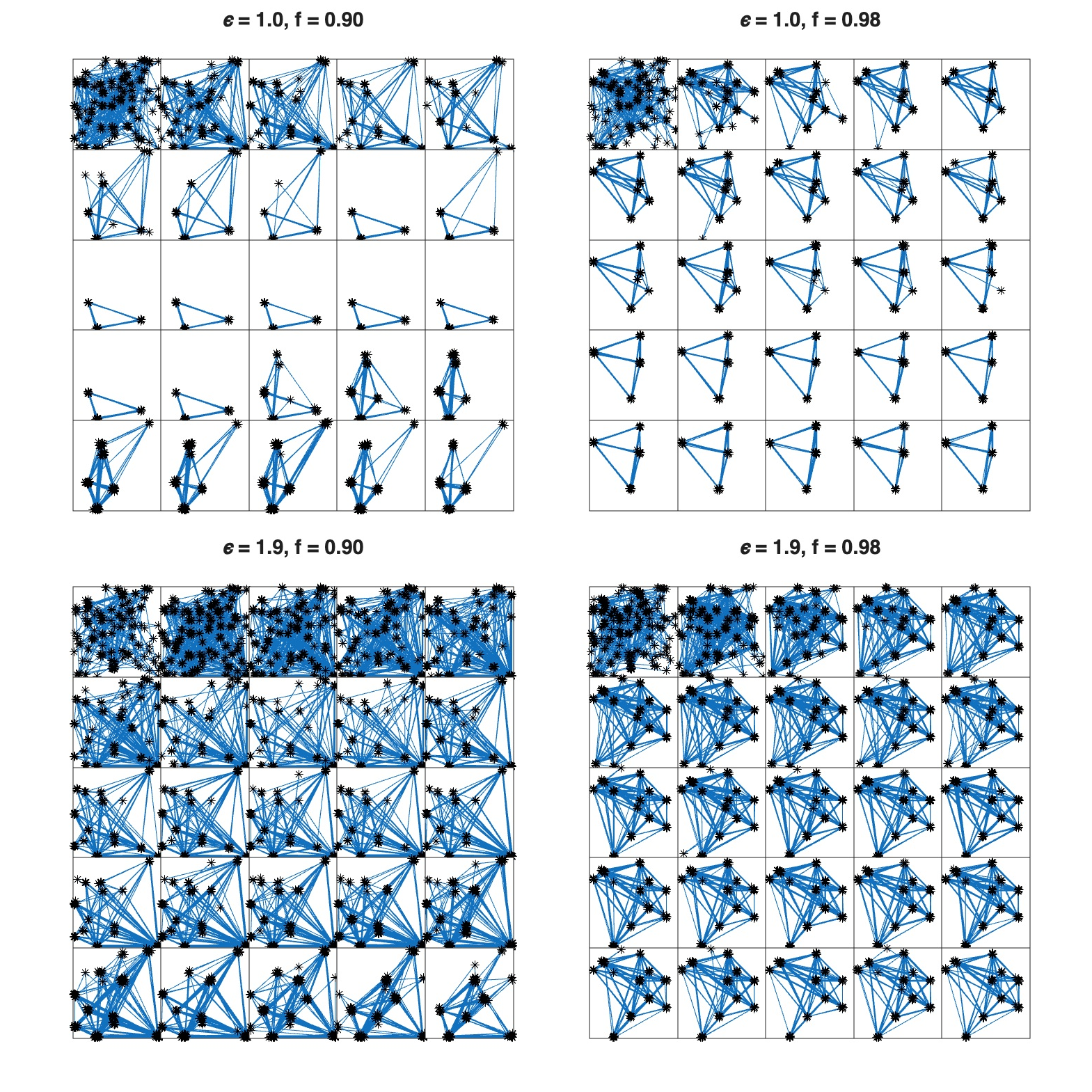}
  \caption[Basic burst origin sequence plots]{Basic burst origin
    sequence plots. Evolution of burst origins during development for
    given $\epsilon$ and $f$. The 25 subgraphs in each plot represents
    large-scale developmental time in row-major order (i.e., top left
    the start of development and bottom right the end). Within each
    subgraph, the $100 \times 100$ neuron culture is displayed with
    burst origin locations during that time period indicated by black
    asterisks ($\ast$) and with each origin connected to the next with
    a solid blue line segment, indicating the small-scale temporal
    sequence of origins from one burst to the next. Burst counts:
    $(\epsilon, f) = (1.9, 0.90)$: 14,880 origins, 596
    origins/subgraph (576 for last subgraph);
    $(\epsilon, f) = (1.9, 0.98)$: 16,815 origins, 673
    origins/subgraph (663 for last subgraph);
    $(\epsilon, f) = (1.0, 0.90)$: 9,446 origins, 378 origins/subgraph
    (374 for last subgraph); $(\epsilon, f) = (1.0, 0.98)$: 9,034
    origins, 362 origins/subgraph (346 for last subgraph).}
  \label{fig:origin-evolution}
\end{figure*}

Figure~\ref{fig:origin-evolution} displays the burst origin locations,
sequences, and the evolution of sequence patterns during development
for each of the four simulations. In all cases, origin locations
tended to be more numerous early in development and later became
restricted to a small subset; this is more pronounced for simulations
with more inhibitory cells. These later sets of origin locations were
stable over long durations --- many thousands of bursts --- though (as
is particularly noticeable for the $\epsilon=1.0$Hz simulations)
specific origins might disappear or appear over long periods of time.

\subsection{Burst and Burst Sequence Probabilities}

%
\begin{figure*}
  \centering
  \includegraphics[width=\textwidth]{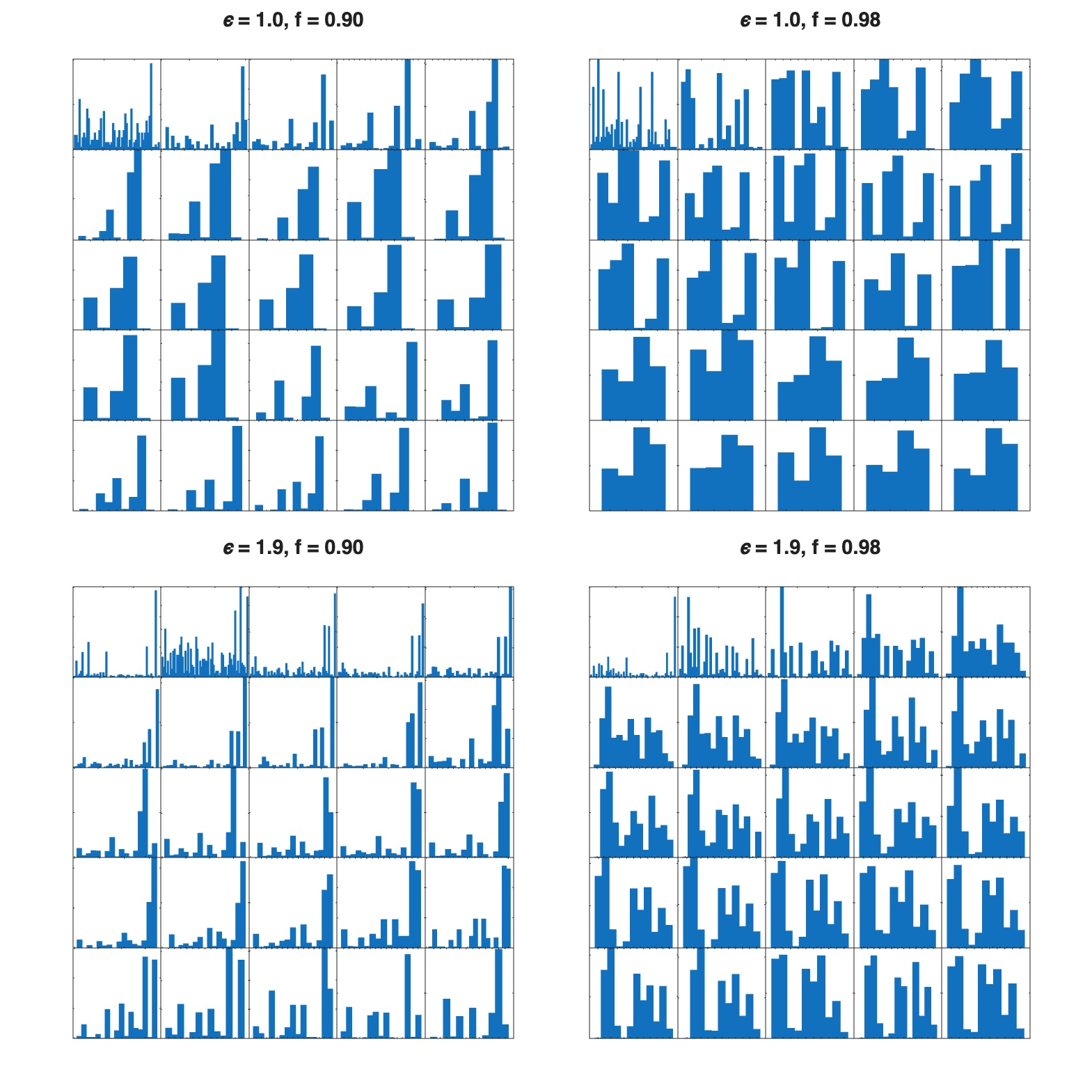}
  \caption{Burst origins histograms during development for given
    $\epsilon$ and $f$. Each of the 25 subgraphs in each plot displays
    a histogram for the corresponding subgraph from
    figure~\protect\ref{fig:origin-evolution}. The histograms are
    normalized as probabilities, so y-axes represent relative
    probability and are formally comparable across subgraphs and
    simulations despite varying numbers of total bursts.  All
    histograms tested non-uniform.}
  \label{fig:origin-histogram}
\end{figure*}

To the extent that burst origins were restricted to a small number of
locations, figure~\ref{fig:origin-evolution} does not clearly show
whether all locations were equally likely or if some were preferred.
Figure~\ref{fig:origin-histogram} presents burst origin histograms for
the four simulations and corresponding subgraphs. These clearly show
the rapid reduction in number of regions in which bursts originate as
development progressed. In almost all burst sequences, the
distribution of bursts across the different origins was significantly
non-uniform by $\chi^2$ test ($p<0.05$): some origins were more
common, though even origins that were less common were generally not
rare.

\begin{figure*}
  \centering
  \includegraphics[width=\textwidth]{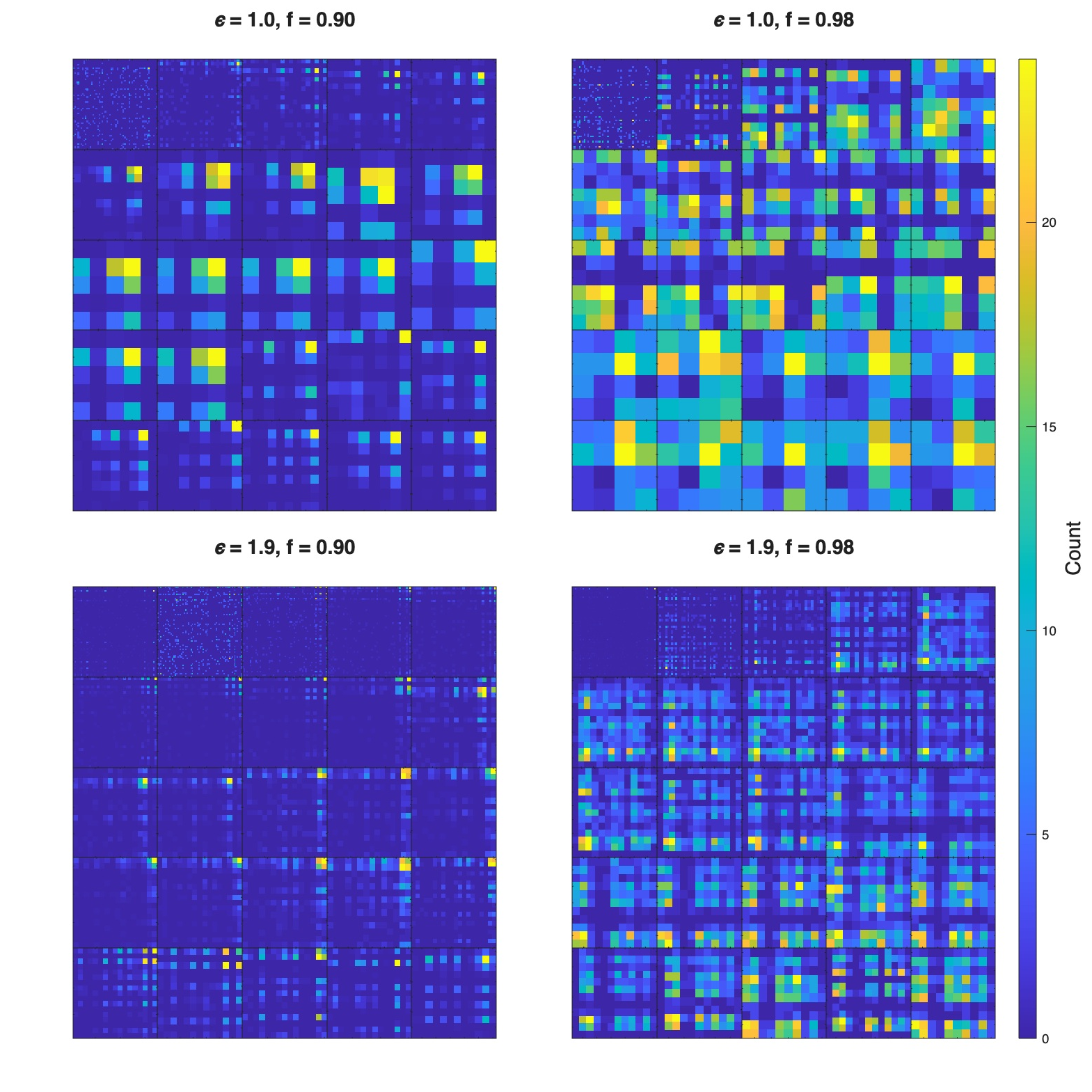}
  \caption{Sequential correlations for burst origins during
    development for given $\epsilon$ and $f$. Each of the 25 subgraphs
    in each plot displays $(\ell_i, \ell_{i+1})$ for the corresponding
    subgraph from figure~\protect\ref{fig:origin-evolution}.}
  \label{fig:origin-markov}
\end{figure*}

Figure~\ref{fig:origin-markov} presents sequential correlations (joint
histograms) between each burst origin tile ($\ell_i$) and the
subsequent burst's origin tile ($\ell_{i+1}$) for corresponding
subgraphs. This aims to test whether the spatial location of one burst
was predictive of the next, beyond what would be expected from their
\emph{a priori} independent marginal frequencies.  If burst origins
were governed by a deterministic sequence, we would expect to see
distinct off-diagonal structure representing specific transition rules
or underlying dynamics. Instead, we observe that bursts in the most
common origin areas were simply most frequently succeeded by another
burst in the same or other common areas, proportional to those areas'
overall likelihoods. Furthermore, the joint histograms are
approximately symmetric across the diagonal, meaning a transition from
region A to region B is roughly as likely as a transition from B to A
(with minor asymmetries likely stemming from the arbitrary fixed-tile
discretization). Because the transition probabilities appear to merely
reflect the product of the independent marginal probabilities without
complex sequential structure, these results strongly support the
assertion that the spatial sequence of burst origins was
indistinguishable from random.

\subsection{Burst Origins and Network Drive}


\begin{figure*}
  \centering
  \includegraphics[width=\textwidth]{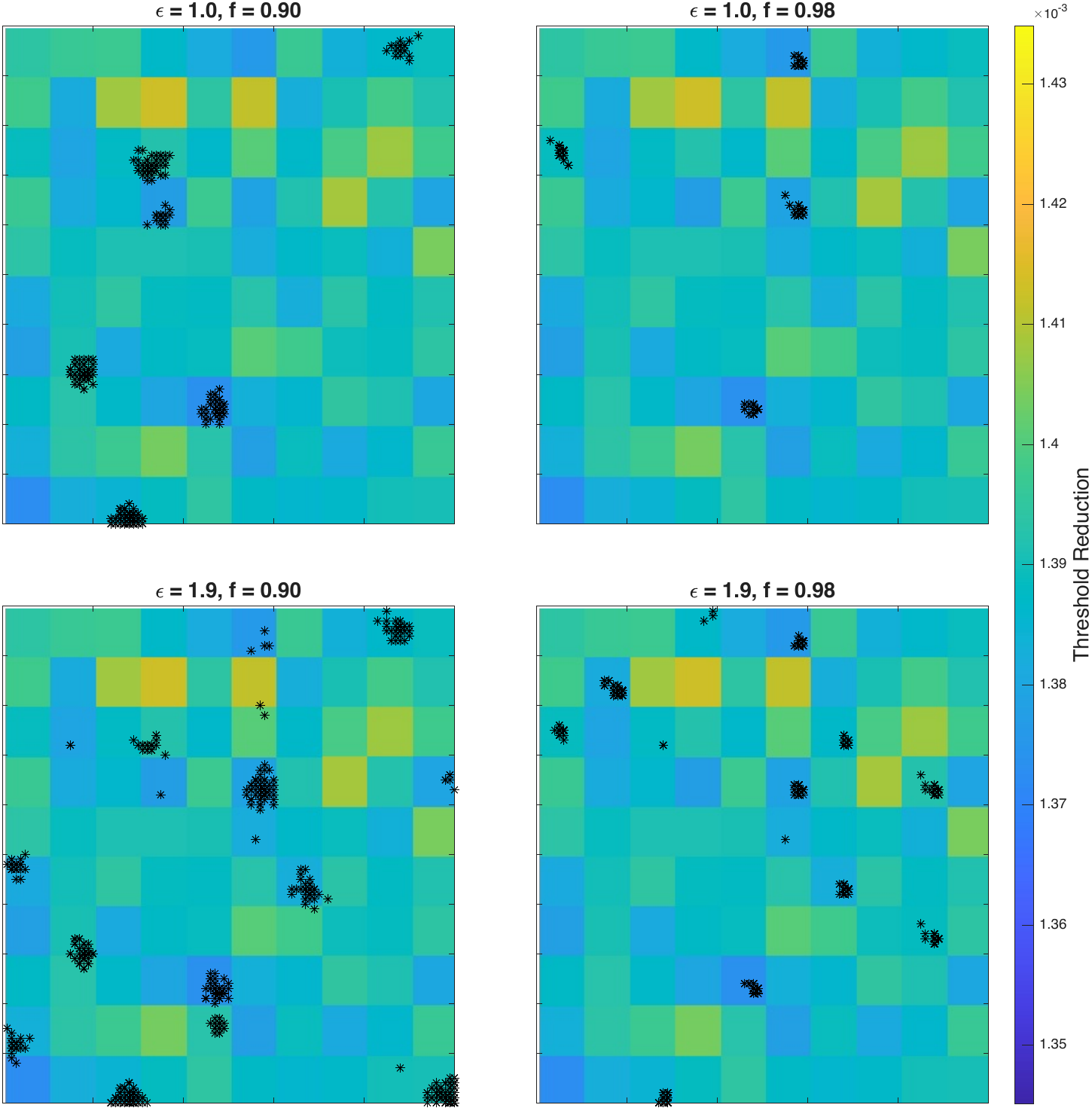}
  \caption{Plots of burst origins and average endogenously active
    neuron lowered threshold for given $\epsilon$ and $f$. Origins are
    for the bursts in the bottom row of the corresponding plots in
    figure~\protect\ref{fig:origin-evolution}. Each $10 \times 10$
    tile of neurons in the network is represented by a color
    indicating the average amount that the 10 endogenously active
    neurons' threshold is lowered; more yellow indicates greater
    average lowered threshold and thus greater average spiking rate
    for those cells.}
  \label{fig:network-drive}
\end{figure*}


\begin{table}
  \caption{Spearman's rank correlation computed to assess the
    relationship between average threshold reduction in each
    $10 \times 10$ tile of neurons and number of burst origins in that
    tile.}
  \label{tab:originDriveCorrel}
  \centering
  \begin{tabular}{|l|l|l|}\hline
    $(\epsilon, f)$ & $\rho$ & $p$ \\ \hline
    $(1.9, 0.90)$ & -0.24 & 0.016 \\
    $(1.9, 0.98)$ & -0.31 & 0.002 \\
    $(1.0, 0.90)$ & -0.13 & 0.193 \\
    $(1.0, 0.98)$ & -0.25 & 0.010 \\
    \hline
  \end{tabular}
\end{table}

We tested the hypothesis that the non-uniformity of burst origin
location could be a result of more active endogenously active neurons
by comparing it to the average network drive nearby.
Figure~\ref{fig:network-drive} relates burst origins near the end of
the simulations (when development became stationary) with the average
amount that the 10 endogenously active neurons in each $10 \times 10$
tile of neurons had their spiking thresholds lowered. In the absence
of synaptic input, a more greatly lowered threshold produces a higher
average (noise driven) endogenous spiking rate. Thus, this is a
measure of the average drive in that tile of neurons and the plots are
a visual indication of whether bursts tend to originate in tiles with
high levels of endogenous activity. By examination, burst origins are
clearly not confined to the most active tiles. This was tested by
computing Spearman's rank correlation between burst count and lowered
threshold for each tile, which is shown in
table~\ref{tab:originDriveCorrel}. One can see that there is a
statistically significant, weak \emph{negative} correlation between
number of burst origins in a tile and network drive for that tile for
three of the four simulations. It is important to note that while
these Spearman correlations are highly statistically significant
(likely due to the very large sample size $N$ of detected bursts), the
absolute effect sizes ($|\rho| \le 0.31$) are weak. This suggests that
while a relationship exists, it is not the primary driver of origin
locations.

\subsection{Burst Origins and Network Edges}


\begin{figure*}
  \centering
  \includegraphics[width=\textwidth]{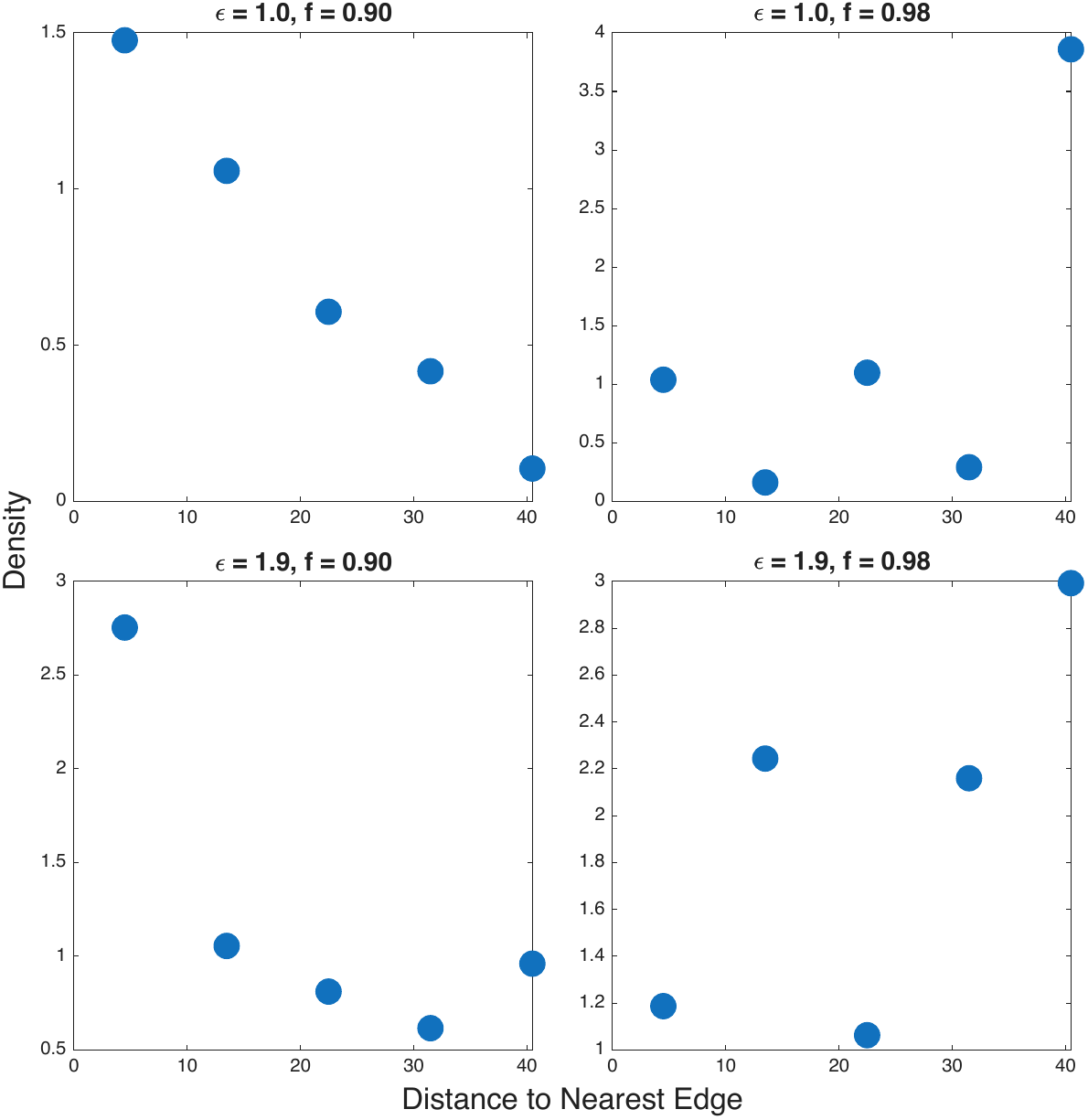}
  \caption{Plots of burst origin densities versus distance to nearest
    network edge for given $\epsilon$ and $f$. All burst origins used
    for each simulation..}
  \label{fig:origin-densities}
\end{figure*}

We analyzed whether burst origin locations could be explained based on
our simulation's boundary conditions by plotting the density of
origins versus distance to nearest network edge in
figure~\ref{fig:origin-densities} . We used a bin size of 5 to roughly
correspond to size of the $10 \times 10$ tiles (and after empirical
exploration of the impact of bin size on clarity of visualization). We
also used all bursts, not just those corresponding to stationary
behavior, as the latter correspond to such a small number of clusters
that it would preclude any reasonable conclusion.

With those provisos, we can see that edges do appear to influence the
density of burst origins, but only for the pair of simulations with
10\% inhibitory neurons. Note that previous results
\citep{kawasaki_simple_2014} demonstrated that the radii of
connectivity for all neurons was less than 3; so only the first bin in
any of these plots could include direct edge effects. In the case of
$(\epsilon, f) = (1.0, 0.90)$, the edge influence on origin location
during development clearly extends well beyond those neurons whose
connectivity is directly, or even indirectly, affected by boundary
conditions.

\section{Discussion}

This paper presented results from closed-loop simulations of
neuron-activity-driven network development, capturing the full
spatiotemporal spiking activity of all neurons. This work used the
simplest neuron/synapse models and network development rules that
duplicate the salient features of \textit{in vitro} burst development
\citep{kawasaki_simple_2014}---leaky integrators with synapses that
include facilitation and resource depletion. While our previous
investigations using this model were confined to analyzing macroscopic
temporal behaviors, the spike-level spatial resolution here enabled us
to examine how spatially localized spiking activity and wave
propagation evolved along time towards stationarity. We also examined
the dependence of that activity on two parameters: (1) the fraction of
inhibitory neurons and (2) the average spiking rate at which neurite
outgrown/retraction was stable.

From these simulations, we can draw a number of conclusions, detailed
in the following subsections, regarding the robustness of the bursting
behavior, the nature of bursting as a network phenomenon, the
spatiotemporal patterns of individual bursts, the evolution of
individual bursts and burst origination as network connectivity
changes, and the relationship between burst origination and local
excitatory drive within the network.

\subsection{Burst development only requires a simple growth model;
  there is no need to fine tune model parameters}

Consistent with previous findings \citep{kawasaki_simple_2014},
bursting develops robustly over a wide range of parameters without
needing precisely tuned growth models. Importantly, whereas our prior
work \citep{kawasaki_simple_2014} was restricted to analyzing
macroscopic time-domain network data, the current study resolves
individual spatiotemporal spikes. This allows us to align with the
macroscopic traveling waves and boundary effects observed by
\citet{gritsun_growth_2012}, while demonstrating these phenomena
emerge naturally from simple distance-based connectivity constraints
\citep{lonardoni_recurrently_2017} without resorting to the
complicated growth algorithms or detailed cell types used in other
studies.

\subsection{Bursting itself is a network phenomenon}

As the individual neuron model used here is not capable of intrinsic
bursting, the rapid firing of individual neurons as they contribute to
wave propagation can only be a product of synaptic input from other
neurons participating in the burst. The decline in spiking that
produces a trailing edge for a wave must result from synaptic
resources being exhausted.  As a result, even in the case of multiple
overlapping bursts, the wavefronts do not continue ``through'' each
other like simple wave propagation, instead combining as they move
through the neurons that have not yet participated in either burst.

This distinction between network bursting and intrinsic neural behavior can
be seen in Figure~\ref{fig:burst-wavefront}, which not only shows a
typical burst, but also reveals the low-level background spiking from
endogenously active neurons before and after the wavefront passes by.
This is possible because endogenous activity does not require synaptic input
(and therefore no synaptic resources).

This is consistent with a broad theoretical literature relating to the
general properties of networks of coupled oscillators
\citep{ermentrout_traveling_2001}, and confirmation of the
applicability of this theory in neural simulations
\citep{sakuma_simulation_2016, senk_conditions_2020}, \emph{in vitro}
\citep{maeda_mechanisms_1995}, and \emph{in vivo}
\citep{sato_traveling_2012, okujeni_mesoscale_2017,
muller_cortical_2018, davis_spontaneous_2021}. Our results seek to
elicit the dependence of bursting characteristics on network
development by looking at when during development bursting starts and how
burst properties change as network connectivity changes. This builds
on previous results that bursts begin soon after connections form
between neurons, that shortly thereafter connectivity patterns for
excitatory and inhibitory neurons diverge, and that burst duration
decreases as development progresses \citep{kawasaki_simple_2014}.

\subsection{Each burst originates at a single point and propagates as
  a wave of activity}

All of the bursts observed in simulations were like that shown in
Figure~\ref{fig:burst-wavefront}, originating at single locations and
propagating as traveling wavefronts across the
network. Figures~\ref{fig:burstSpeed} and~\ref{fig:origin-evolution}
illustrate how wavefront propagation speed and burst origin locations,
respectively, evolved during network development. Such network wave
propagation has frequently been suggested as underlying spatiotemporal
network computation, such as in the visual system
\citep{benigno_waves_2023}.

\subsection{As development progresses, wave propagation speed
  increases}

Figure~\ref{fig:burstSpeed} shows that burst wavefront propagation
speed increased during development \citep[consistent with the findings
of][]{maeda_mechanisms_1995}, by eye reaching stationarity by
simulation end and corresponding to connectivity reaching stationarity
in previous results \citep{kawasaki_simple_2014}.  This increase in
propagation speed explains previous simulation and experimental
results \citep{van_pelt_longterm_2004, vanpelt_long-term_2004,
  kawasaki_simple_2014} that showed that early bursts start out with
relatively long duration and low intensity, becoming briefer and more
intense over time. Previous simulation data indicated the number of
spikes produced did not vary much on average; the current results
indicate that burst duration is a function of propagation speed:
faster propagation corresponds to shorter burst duration. This inverse
relationship arises naturally from the model's resource exhaustion
mechanics. Traveling waves propagate by recruiting and rapidly
consuming available excitable network resources (i.e., through
synaptic facilitation and subsequent depression). A faster-moving
wavefront consumes the culture's finite resources more
rapidly. Because the network cannot sustain high-frequency firing once
local resources are depleted, this faster ``burn'' through the culture
guarantees a shorter overall burst duration. However, regardless of
propagation speed, the width of the wavefront (duration of bursting by
any single neuron) does not appear to change during development and,
since bursts invariably propagate across the entire network, the total
number of spikes within a burst is still constant.

Initial wave propagation speed is strongly influenced by the target
firing rate ($\epsilon$), which drives early connectivity growth. A
higher $\epsilon$ accelerates connectivity formation, resulting in
more frequent, longer-duration bursts early in development
\citep{kawasaki_simple_2014} that initially propagate more slowly but
rapidly increase in speed. Conversely, the final steady-state
propagation speed depends primarily on the fraction of excitatory
cells ($f$), with increased inhibition driving faster waves due to
broader excitatory connectivity \citep{kawasaki_simple_2014}.

A fair amount of variation is seen in the propagation speed computed
for individual bursts at about the same stage in development. Due to
the lack of substantive correlation between propagation speed and
burst origin (table~\ref{tab:speedLocCorrel}), we hypothesize that
this could be ascribed to the analysis algorithm, which only used the
most active cells for each 10ms of a burst.

\subsection{Burst Origins are emergent network phenomena}

Our results show that early in development there are many burst origin
locations, but this number decreases until stationarity
(Figure~\ref{fig:origin-evolution}). We suggest this is consistent
with network connectivity being maximally localized early on
(beginning as 4-connected in these simulations) and becoming broader
as development progresses \citep{kawasaki_simple_2014}. This reduction
of origins occurs without cell death or connection pruning,
demonstrating that even relatively localized connectivity can restrict
burst initiation to a bare handful of functional communities
(fCOMs). A higher target firing rate ($\epsilon$) produces more
origins, while greater inhibition (smaller $f$) slows the evolution
toward the final set of fCOMs (Figures~\ref{fig:origin-evolution}
and~\ref{fig:origin-histogram}).

Given that bursts originate at only certain locations, we investigated
their spatial and temporal independence. Sequential correlations
(Figure~\ref{fig:origin-markov}) were indistinguishable from random,
indicating the origin point for any individual burst does not depend
on preceding bursts.

One potential limitation of our model is the regular $10 \times10$
tiling pattern used for endogenously active (EA) cells, as well as the
finite bounds of the $100 \times 100$ grid. While a rigid grid could
theoretically introduce geometric biases, our results indicate that
burst origins (fCOMs) actually exhibit a weak \emph{negative}
correlation with the layout of endogenous drive
(Figure~\ref{fig:network-drive},
table~\ref{tab:originDriveCorrel}). Because fCOMs do not align with
the most active tiles, the tiling pattern is unlikely to be the
primary driver of origin clustering. Nonetheless, future work should
employ controlled randomization in the locations of EA cells to
definitively rule out tiling artifacts.

A central question is why certain locations correspond to fCOMs while
others do not. If burst origination were purely driven by background
activity, we would expect origins to align with the tiles possessing
the most EA cells. Because they exhibit only a weak negative
correlation instead, burst initiation and the existence of fCOMs must
be emergent structural properties. Mechanistically, this weak negative
correlation may arise from local resource reduction. Highly active
regions constantly deplete their local synaptic resources through
greater background firing, meaning they rarely possess the
``accumulated synaptic potential'' required to trigger a burst. Bursts
therefore more likely originate in nearby, relatively quiet regions
where synaptic resources are more fully recovered.

This sequence of bursting bridges behavioral stochasticity and
structural assembly theory. It follows three phases: a burst produces
a propagating wavefront that exhausts neuronal resources, a highly
variable recovery period follows featuring only background endogenous
activity, and a new burst is eventually triggered. Unlike hard-wired
structural feedforward chains, the ``feedforward'' propagation of
traveling waves in our model emerges \emph{functionally} from resource
exhaustion within a symmetrically recurrent medium. While the timing
of a burst and the ``selection'' of its origin are functionally
stochastic, the highly restricted set of possible origin locations
must be governed primarily by specific structural connectivity
properties \citep{Tetko_2001, Iglesias_2010, Poli_2015}. By showing
that random, functional spatiotemporal triggers are reliably channeled
by a restricted set of structural connectivity patterns to ignite
macroscopic waves, our model bridges classical cell assembly theory
\citep{Palm_1982, Palm_2014} with contemporary analyses of functional
connectivity and traveling waves \citep{Poli_2015, Engel_2021,
  Yuste_2024, Cao_2024}.

\section{Future Work}

This paper establishes a number of activities of developing neural
networks that emerge from the network, rather than the individual
neurons. However, particularly given the localized nature of many of
the events described here (such as burst initiation), their true
interpretation must lie at the mesoscale. In particular, the existence
of a small number of origin locations in the face of a lack of strong
correlation with network drive implies that local connectivity
properties must dictate fCOMs. We hypothesize that certain mesoscale
structures, such as specific network motifs or highly connected hubs,
may serve this role, but future graph theoretic work will be needed to
test this.

Burst also originate at particular points in time. Our results have
demonstrated that neurons and synapses recover from the aftereffects
of a burst and then some variable and significant amount of time
passes before the next. In other words, a burst is not produced just
as soon as the network (or the existing fCOMs) is capable. So, it must
be the case that there is some pattern of triggering activity that
kicks off a burst, and this pattern is somewhat uncommon. What is that
pattern and in what way does it differ from patterns that don't
trigger bursts? How do such patterns interact with or are shaped by
the underlying fCOM structural properties?

We must also recognize that the neurite outgrowth and network
formation simulated here represents just one phase of biological
neural network development. Subsequent to this first stage of
synaptogenesis, where connections are over-produced, comes a
refinement stage where mechanisms such as spike timing dependent
plasticity (STDP) strengthen, weaken, and prune synapses. Just as we
ask questions about the interplay of neuron (microscale), connectivity patterns
(mesoscale), and network (macroscale) for bursting here, STDP
simulation would allow us to examine how network architectures and
behaviors change as connectivity is tuned---both endogenously from the
bursts themselves and due to functional considerations in the presence
of external inputs, outputs, and feedback.

\section{Acknowledgments}

The authors thank Fumitaka Kawasaki, Navee Kaur,
Kate Sprague, and Tom Wong for their helpful discussions and software
development work. MS thanks the Fulbright program of the US Dept. of
State and the Czech Republic Fulbright Commission for their support
and the Institute of Physiology of the Czech Academy of Sciences for
their hospitality and discussions related to this work.

\section{Declaration of generative AI and AI-assisted technologies in
  the manuscript preparation process}

This paper has been proofread and edited by a language model (AI), and
the authors have read through and revised the resulting content to
ensure it accurately reflects the original intent and assume full
responsibility for the content of the published article.

\bibliographystyle{elsarticle-harv}

\bibliography{citations}

\end{document}